\documentclass[amsmath,aps,prl,twocolumn]{revtex4}
\usepackage{amsthm,amsfonts,graphicx,verbatim,color,cancel,amsmath,caption,subcaption}%,subcaption}
\newcommand{\mathsym}[1]{{}}
\newcommand{\unicode}[1]{{}}

\newcommand{\bea}{\begin{eqnarray}}
\newcommand{\eea}{\end{eqnarray}}

\begin{document}

\title{Superconducting pairing in resonant inelastic X-ray scattering}

\author{Yifei Shi$^{1}$, David Benjamin$^{2}$, Eugene Demler$^{2}$ and Israel Klich$^{1}$}

\affiliation{$^{1}$ Department of Physics, University of Virginia, Charlottesville, VA 22904,
USA\\ $^{2}$ Physics Department, Harvard University, Cambridge, Massachusetts 02138, USA}
\begin{abstract}
We develop a method to study the effect of the superconducting transition on resonant inelastic X-ray scattering (RIXS) signal in superconductors with an order parameter with an arbitrary symmetry within a quasiparticle approach. As an example, we compare the direct RIXS signal below and above the superconducting transition for $p$-wave type order parameters. For a $p$-wave order parameter with a nodal line, we show that, counterintuitively, the effect of the gap is most noticeable for momentum transfers in the nodal direction. This phenomenon may be naturally explained as a type of nesting effect.
\end{abstract}
\maketitle

\section{Introduction}
The description of many-body systems is usually only practical in terms of simplified low energy theories. Such theories are indispensable and describe a large variety of measurements such as conductance and magnetic response. However, measurements based on scattering techniques often probe wider energy scales. Indeed, powerful probes such as resonant inelastic X-ray scattering (RIXS), are allowing unprecedented access to a wide range of excitations in superconducting and magnetic systems.  In particular, the superconducting gap scale is tiny in comparison with band energies of most materials and often below the experimental energy resolution scale. It is therefore of interest to ask to which extent details of low energy theories, such as the gap function are observable through RIXS. 
Recently, it was suggested that RIXS can distinguish between different phases of the order parameter \cite{marra2013resonant,Marra:2016aa}. This dependence is studied through the dynamical structure factor which is shown to discriminate between singlet and triplet pairing. The structure factor itself is related to the RIXS signal only in the limit of ultra short core hole life time, for which a more elaborate treatment is needed \cite{jia2016using}.

Here, we set out to examine the effect of superconducting pairing on the RIXS mechanism within a simple mean field BCS picture, which includes the effect of core hole potential and goes beyond the ultra short core hole life time approximation which is used to relate RIXS with dynamical structure functions.
We derive a general formula for the RIXS intensity for an arbitrary quadratic Fermi Hamiltonian, with anomalous pairing $\Delta$, as expressed in Eq. \eqref{S integral} together with \eqref{expression for Smn}. 
This result generalizes the quasi-particle approach of \cite{benjamin2014single}, where the computation of RIXS spectra was performed using a model of non-interacting quasiparticles but including an interaction with a positively-charged core hole via exact determinant methods. 
This formalism allows us to compute the characteristics of the signal by numerically evaluating \eqref{expression for Smn}. Moreover, the computations can be done for arbitrary band structures using relatively straightforward numerical means. 

As a demonstration of the method, throughout the paper we will concentrate on $p$ wave superconducting states. $p+i p$ superconductors are of great current interest. Such superconductors can support unpaired Majorana fermions at cores of (half quantum) vortices \cite{Alicea2012,HasanKane2010}, and allow for non-Abelian statistics \cite{ivanov2001non,Alicea2010majorana}.  
Remarkably, we find that the RIXS signal is sensitive to the presence of a superconducting gap $\Delta$, even down to a scale where $\Delta$ is quite small (a few percent) compared to the value of band parameters. In particular, going through the superconducting phase transition $\Delta$ acquires a non-zero value and we expect the RIXS spectra to experience a significant change.

Resonant inelastic X-ray scattering is an important technique for the investigation of a large variety of excitations in correlated systems. Its main advantage is the wide range of energy scales to which it is sensitive: from low energy excitations, such as phonons, to charged excitations of several $eV$. Another advantage is that it is a bulk measurement.
The physical mechanism at play in a RIXS experiment is a second-order photon absorption process, involving a shake-up of the system due to an abrupt appearance of a core hole potential. The non-equilibrium process involved may be rather complicated, and thus the interpretation of experimental measurements may not be straightforward.

%In a RIXS process, a core electron absorbs an X-ray photon and is raised to an excited state. This excited state in turn is affected by the complicated dynamics of conduction band electrons, and eventually the core is filled again and a photon is emitted.  

In the process, photons with energy $\omega$ and momentum $\bold{q}$, are scattered, and the outgoing photons have energy $\omega-\Delta\omega$, and momentum $\bold{q+Q}$ (we take $\hbar=1$ throughout). A complete description of the RIXS intensity would require the consideration the full interacting dynamics of the sample, which is too hard to achieve. Below we will start from the standard approach, using from the Kramers-Heisenberg cross section\cite{ament2011resonant}:
\bea\label{KHform}
I(\omega,\bold{k},\bold{k^{'}}) \propto \sum_f \mid \mathcal{F}_{fg} \mid ^2 \times \delta(E_g-E_f+\Delta\omega),
\eea
with 
\bea
\mathcal{F}_{fg}=\sum_{l,n} e^{\text{i}\bold{Q}\cdot \bold{R_n}} \frac{\langle f\mid d_{n} \mid l\rangle \langle l \mid d_{n}^\dagger \mid g \rangle}{E_g+ \omega-E_l+\text{i} \Gamma}. \label{intensity}
\eea
Here, $|f\rangle,|g \rangle$ are the initial and final state, respectively, of the electron band, and $E_{f,g}$ are their energies. The operator $d_n$ creates a quasiparticle in a conduction band at site $\bold{R_n}$. The states $|l \rangle $ are the set of eigenvectors of the intermediate Hamiltonian $H_n=H+V_n$, where the remaining core-hole is interacting with the conduction band through a potential $V_n$. The form of the potential $V_n$ may be arbitrary. 
In this paper we used both the local form $V_n=U_c d_n^\dagger d_n$, describing an on site interaction with a local core hole, as well as $V_n=U_c  d_n^\dagger d_n+U'_c \sum_{|{\bold{R_n}}-{\bold{R_n'}}|=1} d_{n'}^\dagger d_{n'}$, to account for the effect of the coulomb interaction on the neighboring sites. Here $\Gamma$ is the inverse of the core-hole lifetime, which we take a typical value of order $0.1eV$.

It is important to note that the Kramers-Heisenberg formula \eqref{KHform} is incomplete in that it doesn't properly account for the photoelectron-core-hole Coulomb interaction (see e.g. \cite{zaanen19852,kruger2004x}). Here, however, we neglect such effects as we are only interested in the physics involving the band structure. Indeed, these effects (for example, mixing between $L_2$ and $L_3$ absorption edges) are more pronounced in lighter elements, while in heavier elements of interest for high $T_c$ superconductivity as well as the $p$-wave system described here, the $L_2,L_3$ separation in energy is very large (of order $20eV$ for $Cu$ and $130eV$ for $Ru$).

\section{An exact determinant formula using Majorana fermions}
Following \cite{benjamin2014single}, we write the intensity as:
\bea\label{S integral}
I &\propto& \int_{-\infty}^\infty \text{d}s \int_{-\infty}^\infty \text{d}t \int_{-\infty}^\infty \text{d}\tau 
e^{\text{i}\omega (\tau-t) - \text{i}s\Delta \omega - \Gamma(t+\tau)} \nonumber\\
&\times& \sum_{mn} e^{\text{i}\bold{Q}(\bold{R_m-R_n})} S^{mn},
\eea
with
\bea\label{Smn}
S^{mn}=\langle e^{\text{i}H\tau} d_n e^{-\text{i}H_n\tau} d_n^\dagger e^{\text{i}Hs} d_m e^{\text{i}H_mt} d_m^\dagger e^{-\text{i}H(t+s)} \rangle.
\eea
As long as the various stages in the time evolution are governed by quadratic Fermi operators, \eqref{Smn} can be calculated by exact diagonalization methods. Consider fermions on a lattice with $N=L\times L$ sites, governed by a mean field Hamiltonian:
\bea
H &=& \sum_{i,j} h_{ij} d_i^\dagger d_j + \Delta_{ij} d_i d_j + \text{h.c.} 
\label{Hamiltonian}
\eea
To handle arbitrary superconducting pairing $\Delta_{ij}$, we represent the fermion creation and annihilation operators in terms of $2N$ Majorana fermions $c_k$ defined as:
\bea
c_k = \begin{cases}
d_k+d_{k}^\dagger & \small{ {k=1,2,...N}} \\
\text{i}(d_{k-N}^\dagger - d_{k-N}) & \small{ {k=N+1,N+2,...2N}}~,
\end{cases}
\eea
and satisfying the relation $\{ c_i, c_j \} = 2\delta_{ij}$. The Hamiltonian \eqref{Hamiltonian} can be re-expressed in terms of the Majorana fermions as 
\bea\label{H majoranas}
H = \sum_{ij} \mathfrak{h}_{ij} c_i c_j,
\eea
with $\mathfrak{h}$ the antisymmetric matrix:
\bea
\mathfrak{h}= \frac{1}{4}
\begin{pmatrix}
i\text{Im}(h+2\Delta) & i\text{Re}(2\Delta + h) \\
i\text{Re}(2\Delta - h) & i\text{Im}(h-2\Delta)
\end{pmatrix}.
\eea
Traces involving quadratic Hamiltonians of the form $A=\mathfrak{a}_{ij} c_i c_j$ where $\mathfrak{a}_{ij}$ is an anti-symmetric matrix, can be calculated by using the counting statistics formulas presented in, e.g. \cite{klich2014note}. As shown in the appendix, the trace formula
\bea\label{Trace formula}
\text{Tr}(e^{A_1}...~e^{A_n}) = \sqrt{\text{det}(1+e^{4\mathfrak{a}_1}...e^{4\mathfrak{a}_n})},
\eea
leads, in the direct RIXS case, to the three distinct contributions to $S^{mn}$,
\bea
S^{mn} &=& S^{mn}_1+S^{mn}_2+S^{mn}_3.
\eea
The contributions are detailed in the Appendix, but we mention that in the absence of a core hole potential $S_2$ contributes only to the elastic signal, while in the absence of superconducting pairing, $S_3$ vanishes. We note that the sign of the square root in equation \eqref{Trace formula} is determined to be consistent with analyticity of the expression as function of $t,s,\tau$. The first term is given explicitly by:
\small{ 
\bea\label{expression for Smn}
S^{mn}_{1} &=& \sqrt{\text{det}(F)}(\Lambda_{n,m}+\Lambda_{n+N,m+N}-\text{i}\Lambda_{n+N,m}-\text{i}\Lambda_{n,m+N}) \nonumber \\
 & & \times  (\Gamma_{m,n} + \Gamma_{m+N,n+N} - \text{i}\Gamma_{m+N,n} + \text{i} \Gamma_{m,n+N}).
\eea
}
\normalsize
Here $\Lambda_{n,m}, \Gamma_{n,m}$ are elements of the $2N\times2N$ matrices
\bea
\Lambda & = & e^{i \mathfrak{h} s}e^{i \mathfrak{h}_{m}t}e^{i(\tau-t-s)\mathfrak{h}}G^{-1}(1-N_{\beta})e^{-i(\tau-t-s)\mathfrak{h}}e^{-i \mathfrak{h}_{m}t}\nonumber\\
\Gamma & = & e^{i(\tau-t-s)\mathfrak{h}}N_{\beta}F^{-1},
\eea
where  $N_\beta=\frac{1}{1+e^{-4\beta \mathfrak{h}}}$, $K=e^{-4i \mathfrak{h}_{n}\tau}e^{4i \mathfrak{h} s}e^{4i \mathfrak{h}_{m}t}e^{4i(\tau-t-s) \mathfrak{h}}$, $F=1-N_\beta + KN_\beta$, $G=1-N_\beta + N_\beta K$. Here $\mathfrak{h}_{m}$ represent the Hamiltonian with core hole at position $m$ (i.e. $H_{m}=\sum_{ij} (\mathfrak{h}_{m})_{ij} c_i c_j$). We stress that the equations (\ref{S integral},\ref{Smn},\ref{expression for Smn}) are valid for any type of mean field pairing and are the main technical result. We  now turn to apply these for a particular pairing, of $p$ wave form.

\section{Application to a $p+ip$ superconductor}
To be concrete, we take a minimal toy model for a $p$ wave superconductor. We use a two-dimensional, spinless fermionic system, on a square lattice, with superconducting gap $\Delta$. In the Hamiltonian \eqref{Hamiltonian}, we choose band structure parameters sometimes used for Strontium Ruthenate, $Sr_{2}RuO_{4}$. Following \cite{lederer2014suppression}, we choose $h_{ii} = -\mu$, $h_{i,i+\hat{x}} = h_{i,i+\hat{y}} = -t_1$, $h_{i,i\pm\hat{x}\pm\hat{y}} = -t_2$. To get a $p_x+i p_y$ superconducting state, we take, to be concrete,  $\Delta_{i,i+\hat{x}}=\Delta$, $\Delta_{i,i+\hat{y}}=\text{i}\Delta$, with $(\mu, t_1, t_2, \Delta) = (1.15, 0.8, 0.3, 0.05)\varepsilon$, where $\varepsilon\sim0.2eV$ \cite{kontani2008giant}. 
In comparisons with $Sr_{2}RuO_{4}$, the Hamiltonian \eqref{Hamiltonian} is associated with the so-called $\gamma$ band of  the $d_{xy}$ orbitals in the Ruthenate. The signal may also get contributions from additional quasi-1d bands associated with $d_{xz},d_{yz}$ orbitals, with hopping  $\sim\varepsilon$. We have also carried out explicit calculations for the $d_{xz}$ and $d_{yz}$ bands, however we focus here on the $\gamma$ band.

\begin{figure} 
%\begin{center}
\includegraphics[width=0.238\textwidth]{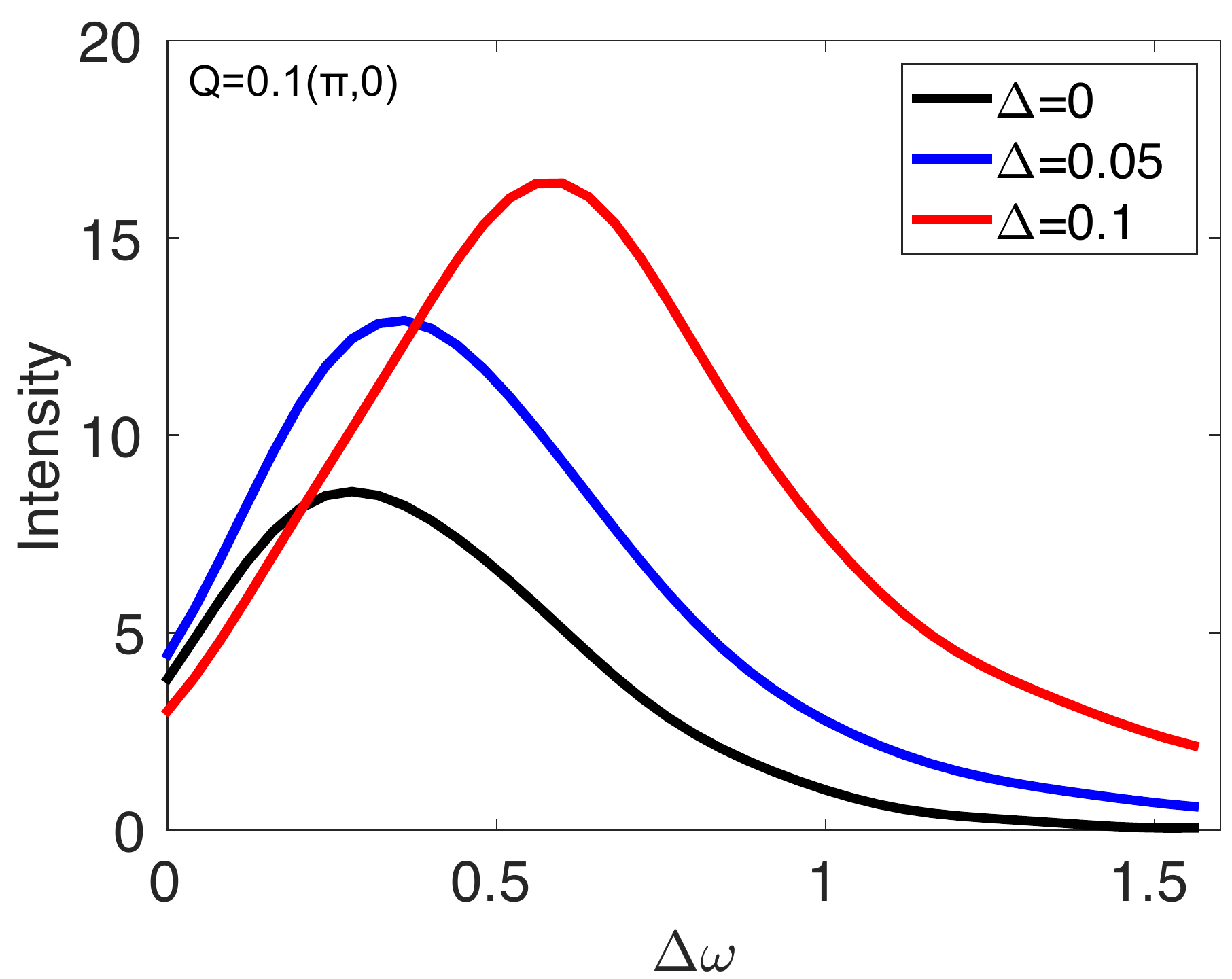}
\includegraphics[width=0.238\textwidth]{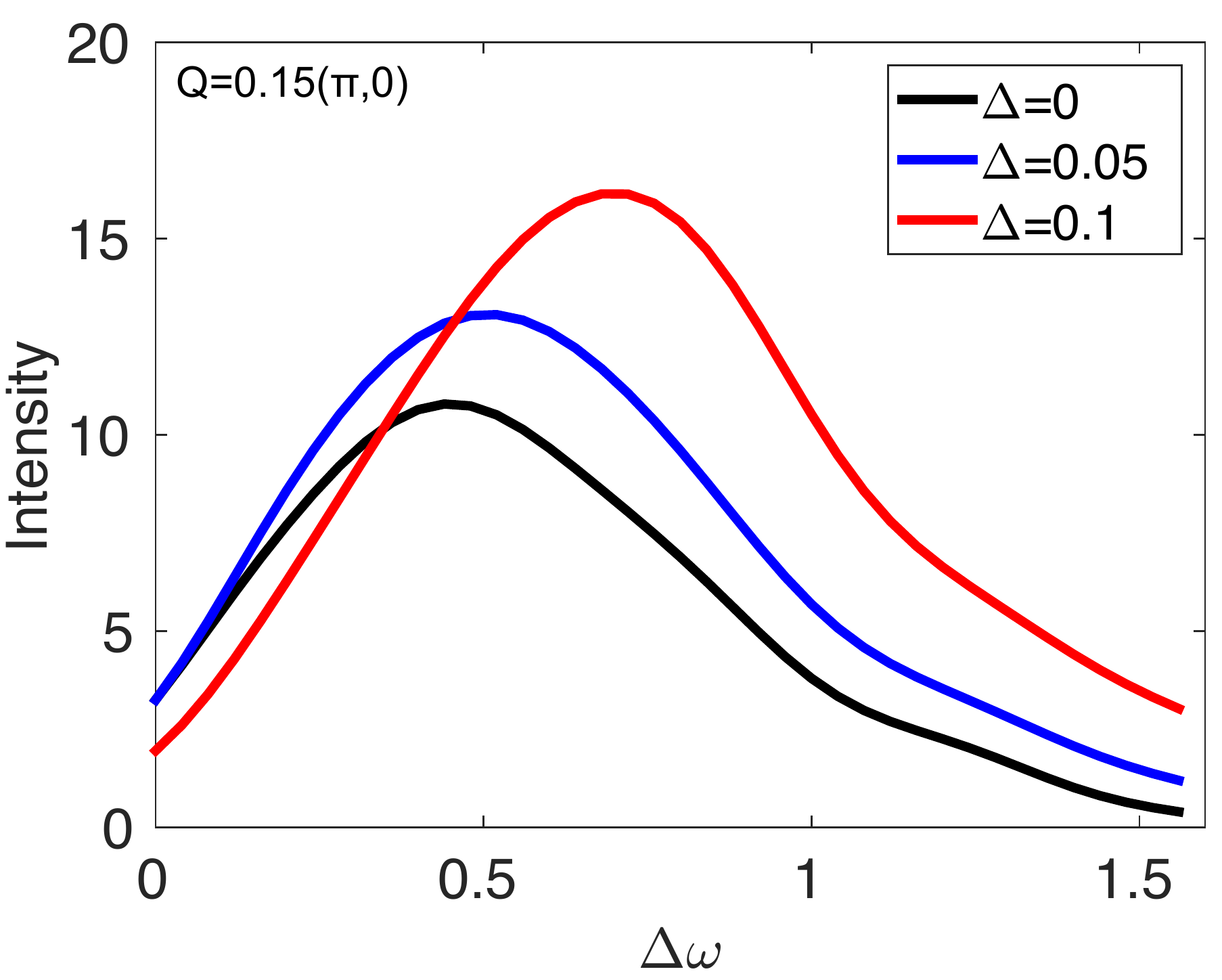}\\ 
\includegraphics[width=0.238\textwidth]{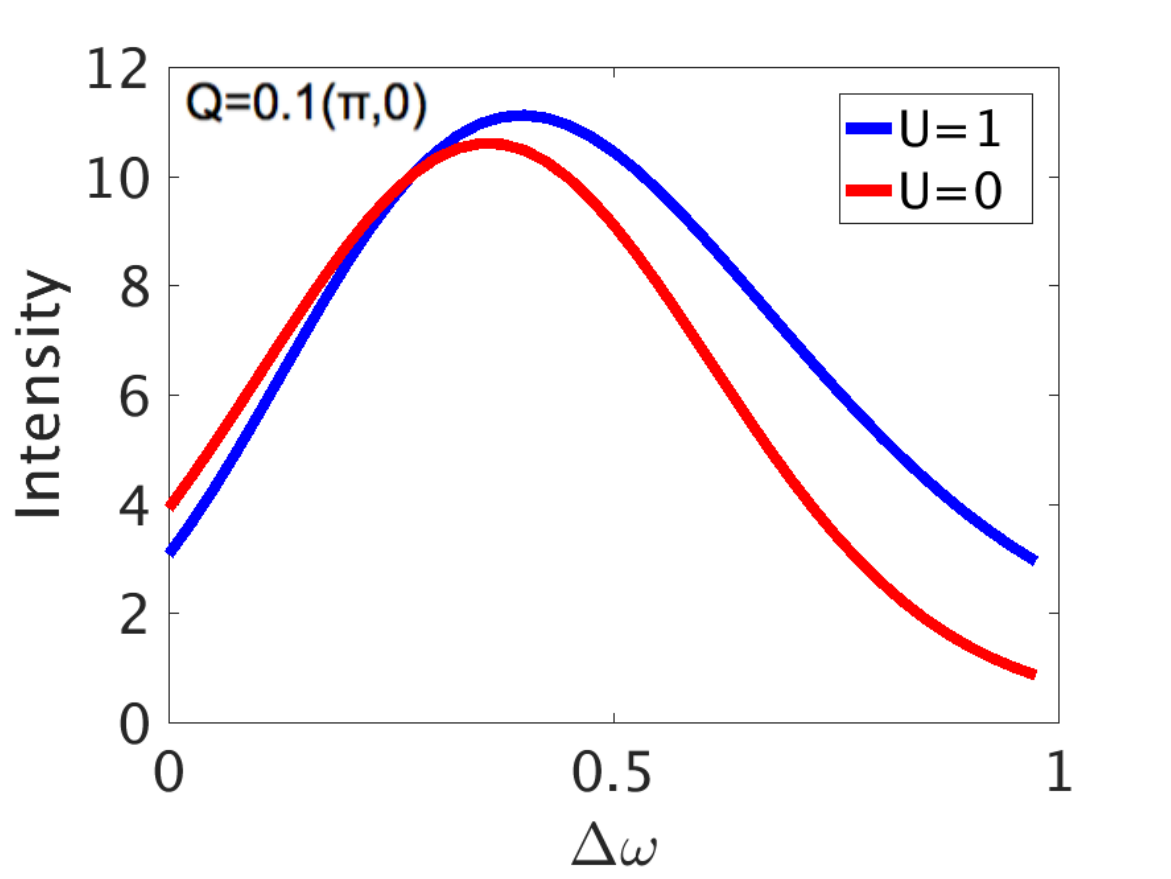}
\includegraphics[width=0.238\textwidth]{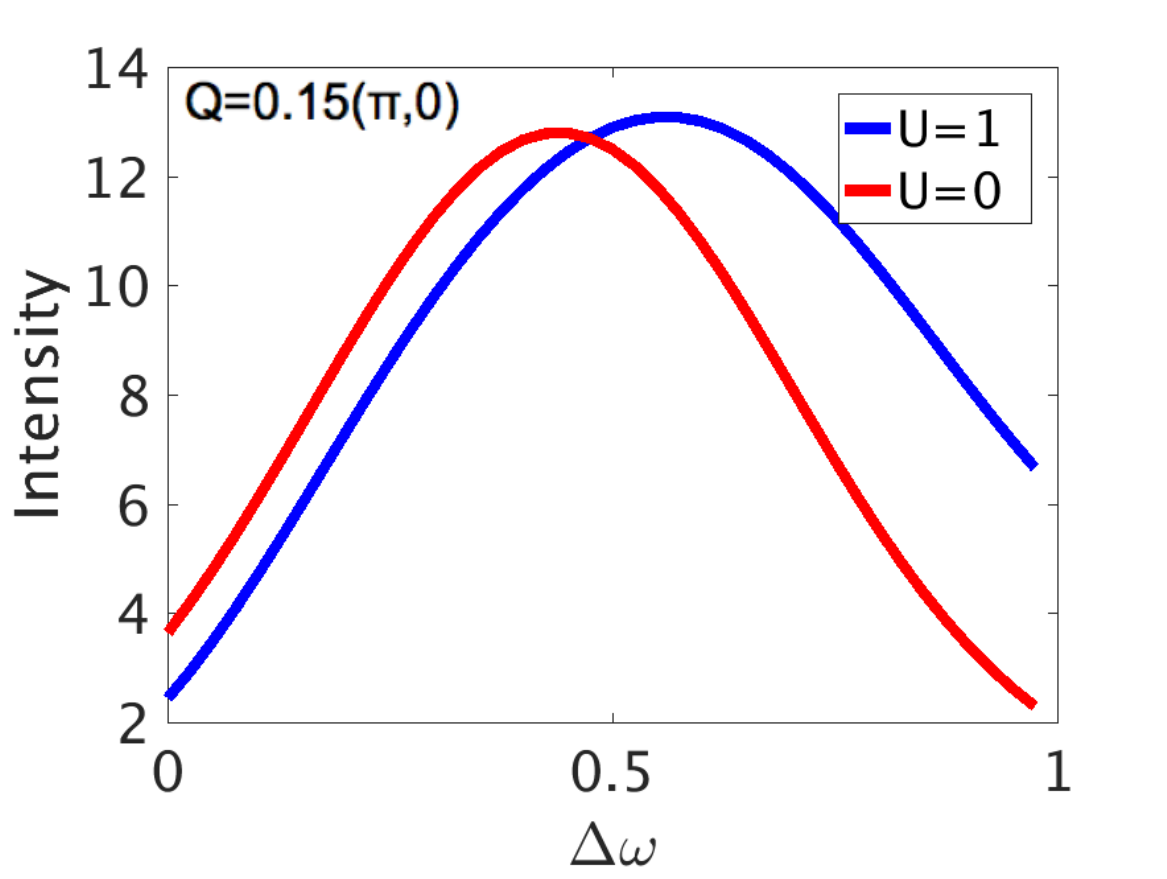}
%{0227_3DIntensity.png}
\caption{Upper panels: RIXS intensity across the transition for a $p_x+i p_y$ superconductor. (Left panel: $\bold{Q}=0.1(\pi,0)$; Right panel: $\bold{Q}=0.15(\pi,0)$), $U_c=1\varepsilon$, for different values of $\Delta$ (in units of $\varepsilon$).  $\Delta$ increases the spectral weight for higher energy exchanges, shifts the peak position, and increases the intensity. Notice the shift of the peak position is not linear in $\Delta$. Lower panels: spectral shift due to core hole potential as compared to $U_c=0$ for $\Delta=0.05\varepsilon$. }%\end{center}
\label{fig:IntensityTransiton}
\end{figure}

To explore the role of the superconducting gap, we calculated the RIXS intensity across the superconducting phase transition using Eq. \eqref{expression for Smn}. As is shown in Fig. \ref{fig:IntensityTransiton}, for $\bold{Q}=0.15(\pi,0),0.1(\pi,0)$, the main effect seems to be the shift of spectral density to higher energies:  the intensity decreases for small energy transfer and increases for large energy transfer, and the shift is not simply proportional to $\Delta$, another observable effect is the increase in intensity at the peak. Calculations were carried out at zero temperature both in the presence and without the gap (We have found that thermal corrections beyond the presence of the gap do not play a significant role in the RIXS signal). 
The lower panels in Fig. \ref{fig:IntensityTransiton} exhibit the spectral shift cause by including a core hole potential in the calculations. Introduction of a core hole tends to shift spectral weight to higher energy exchanges due to the available coulomb potential. This contribution is taken into account exactly in the full formalism developed between Eqs. (\ref{S integral}-\ref{expression for Smn}) and is vital for a full comparison with possible experiments. However, we have found that in most of the calculations a qualitative description of the effects of a finite gap $\Delta$ on the intensity curves works quite well already at $U_c=0$, and will be described below.

The spectral density flow to higher energy can be simply understood by noting that the main contributions to RIXS intensity occur due to the generation of electron-hole pairs where one of them is close to the Fermi level. In the presence of pairing, states close to the Fermi level are unavailable - the incoming photon must first overcome the energy gap, and thus the energy difference in the subsequent electron-hole pair is higher. On the other hand, as we see below, surprisingly, the increase in intensity at a direction $\bold{Q}$ is not directly related to the pairing $\Delta_{\bold{Q}}$ at that wave vector.

\begin{comment}
%%%%%%%%%%%%%%
\begin{figure} 
%\begin{center}
\includegraphics[width=0.4\textwidth]{0225Changeofintensity.png}
\caption{The change of intensity $I_{\Delta=0.05t}-I_{\Delta=0}$, for $\bold{Q}=\pi(q,0)$ at several $q$ values.}
\label{fig:IntensityChange}
%\end{center}
\end{figure}
\end{comment}
%%%%%%%%%%%%

\section{The $U_C=0$ approximation}
As stated above, many interesting differences in the RIXS signal below and above the SC transition can be observed already for small $U_c$. 
In this limit we can compute the RIXS more efficiently using perturbation theory. 
We consider an expansion in terms of $U_c$ for $\mathcal{F}_{fg}$ in \eqref{intensity}. For a simple on-site core hole potential $V$ we write: 
\bea
G&=&(H_{m}-E_{g}+i\Gamma+\omega)^{-1}\nonumber\\
%&=&((G^{(0)})^{-1}+U_{c}(d_{m}^{\dagger}d_{m}))^{-1}\nonumber\\
&\sim& G^{(0)}-U_{c}G^{(0)}(d_{m}^{\dagger}d_{m})G^{(0)}+...\nonumber
%\\ &=&G^{(0)}+G^{(1)}
\eea
where $G^{(0)}=(H-E_{g}+i\Gamma+\omega)^{-1}$, is the propagator with no core-hole. %$G^{(0)}$ and $G^{(1)}$ give the zeroth and first order contribution to the intensity. 
From here on we take only the lowest order contribution, where $U_c=0$. The theory is then exactly solvable in terms of the eigenstates of the static problem, and we can calculate the intensity efficiently. We first solve the energy spectrum by switching to momentum space and writing the Hamiltonian in the standard Bogoliubov-de Gennes form:
\bea
H=\frac{1}{2} \sum_\bold{k}
\begin{bmatrix}
d_\bold{k}^\dagger & d_\bold{k}
\end{bmatrix}
\begin{bmatrix}
\epsilon_k & \Delta_\bold{k} \\
\Delta_\bold{k}^{*} & -\epsilon_\bold{k} 
\end{bmatrix}
\begin{bmatrix}
d_\bold{k} \\
d_{-\bold{k}}^\dagger
\end{bmatrix}
\eea 
where $\epsilon_\bold{k} = -\mu - 2t_1 [\text{cos}(k_x)+\text{cos}(k_y)] - 4t_2 \text{cos}(k_x)\text{cos}(k_y)$, and $\Delta_k = 2\text{i}\Delta [{\sin}(k_x)+{i\sin}(k_y)]$, the Hamiltonian is diagonized by a Bogoliubov transformation:
\bea
d_\bold{k} &=& u_\bold{k}^*b_\bold{k} + v_\bold{k} b_{-\bold{k}}^\dagger \nonumber \\
d_{-\bold{k}}^\dagger &=& -v_\bold{k}^* b_\bold{k} + u_\bold{k} b_{-\bold{k}}^\dagger 
\eea
the energy of the excitation is now $E_\bold{k} = \sqrt{\epsilon_\bold{k}^2+\mid\Delta_\bold{k}\mid^2}$, $\mid u_\bold{k}\mid^2+\mid v_\bold{k} \mid^2 = 1$,  and $\frac{u_\bold{k}}{v_\bold{k}}=\frac{\Delta_\bold{k}}{E_\bold{k}-\epsilon_\bold{k}}$, the ground state is annihilated by all $b_k$s, and $\mathcal{F}_{fg}$ in Eq. \eqref{intensity} is now given explicitly by:
\bea
\label{firstorder}
\mathcal{F}_{fg}^{0} \!=\!\sum_{\bold{k_1,k_2,r}} \!\! e^{\text{i}\bold{r}\cdot(\bold{k_1-k_2+Q})} \frac{v_{\bold{k}_1} u_{\bold{k}_2}}{E_{\bold{k}_2}+\omega+i\Gamma}\langle f |b_{-\bold{k}_1}^\dagger b_{\bold{k}_2}^\dagger | g \rangle  
\eea
\begin{figure}
%\begin{center}
\includegraphics[width=0.4\textwidth]{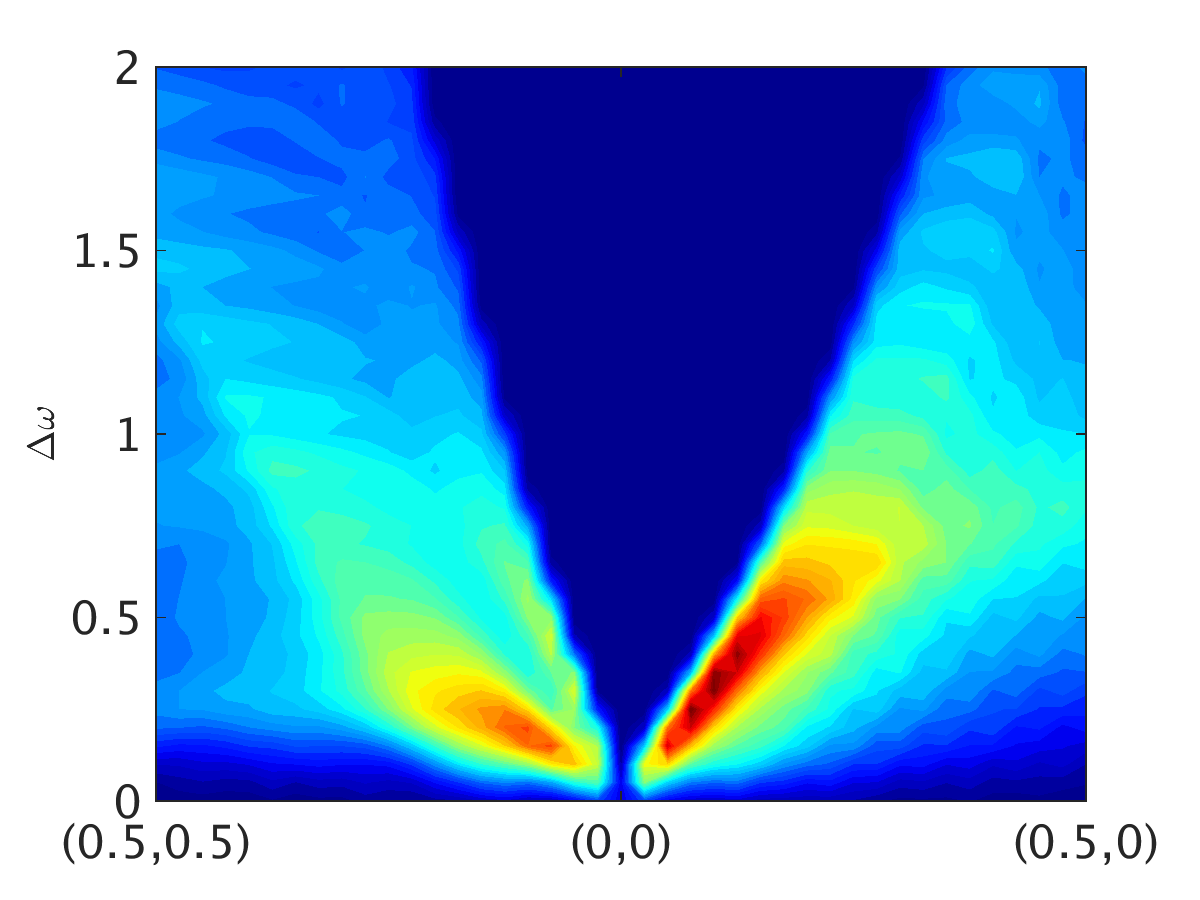}
\includegraphics[width=0.4\textwidth]{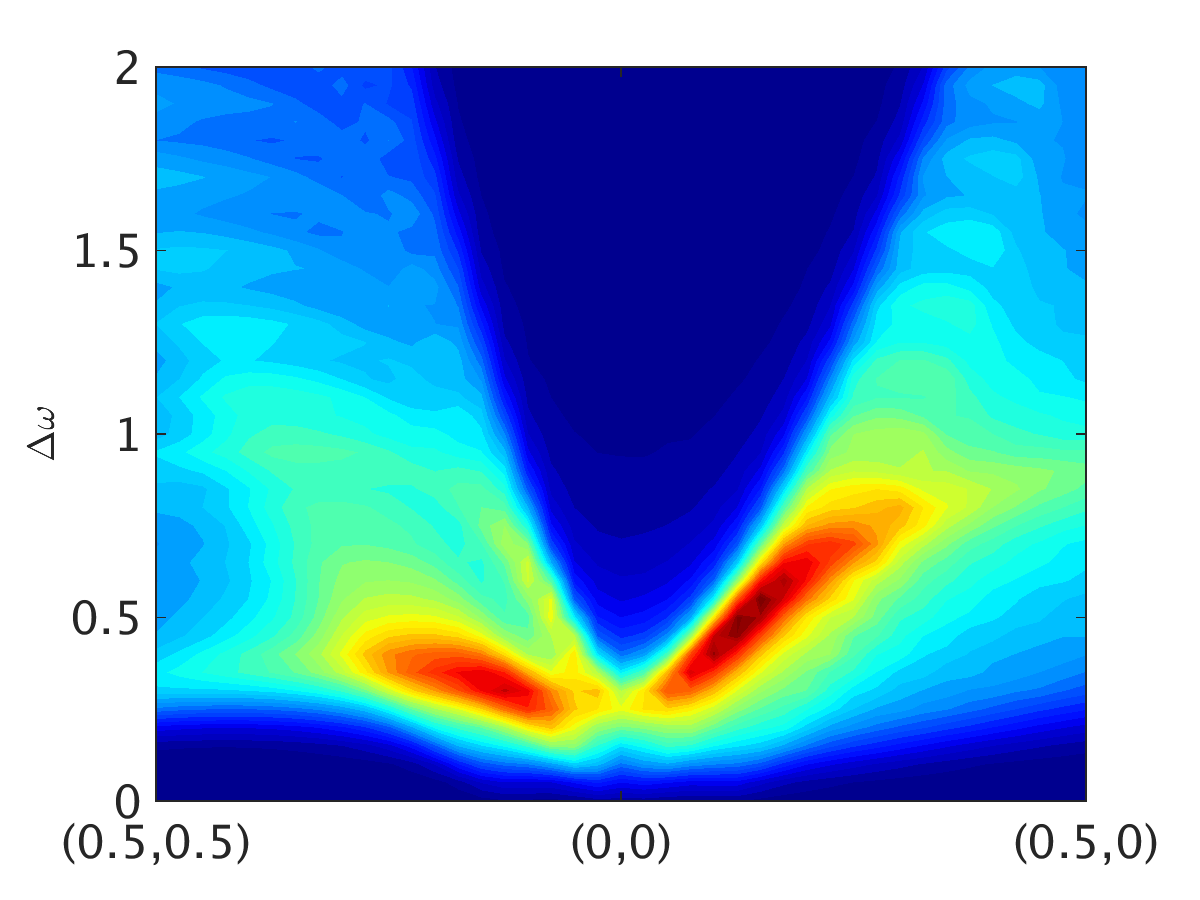}
\caption{RIXS intensity map on $\bold{Q}$, $\Delta\omega$ plane, in the $(11)$ and $(10)$ directions (in units of $\pi$). Upper panel is in the normal phase ($\Delta=0$), lower panel is the superconducting phase ($\Delta=0.05\varepsilon$). The calculation is done using Eq. \eqref{firstorder}.}
%\end{center}
\label{fig:Map}
\end{figure}
\begin{figure}
%\begin{center}
\includegraphics[width=0.4\textwidth]{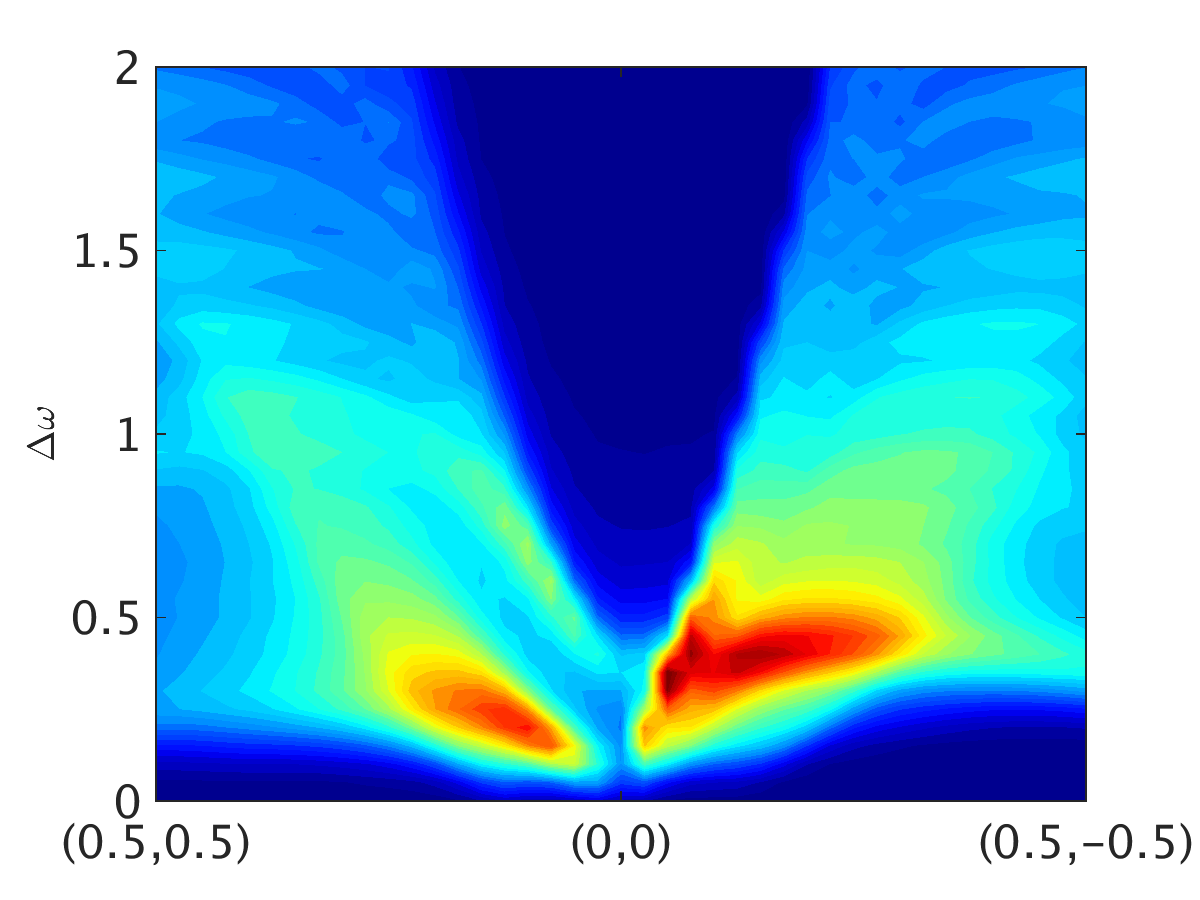}
%\end{center}
\caption{RIXS intensity map for a spinless $p_x+p_y$ type of pairing function ($\Delta_k=i2\Delta(\text{sin}k_x+\text{sin}k_y$)) with $\Delta=0.05\varepsilon$, in the anti-nodal $(1,1)$, and nodal directions $(1,-1)$ for $\bold{Q}$. In the absence of pairing the two directions should have the same intensity, thus the difference comes purely from the presence of the superconducting gap. Surprisingly, the effect is more pronounced on the nodal direction where $\Delta=0$.}
\label{fig:Map_pp}
\end{figure}
%As remarked before, Fig. \ref{fig:IntensityChange}, shows the change in intensity below and above the superconducting transition/
From \eqref{firstorder} we see that in the quasiparticle picture, the contribution to RIXS intensity comes from pairs of quasiparticles with momenta $\bold{k}$ and $\bold{k+Q}$, energies $E_\bold{k}$ and $-E_\bold{k}+\Delta\omega$. When there is no pairing term, these are an electron and a hole, and in the presence of a pairing term, these are the Bogoliubov quasiparticles. Going to the superconducting phase, the energy spectrum becomes $E_k=\sqrt{\epsilon_\bold{k}^2+|\Delta_\bold{k}|^2}$, when $|\epsilon_\bold{k}|\gg |\Delta_\bold{k}|$, we have $E_\bold{k}\sim |\epsilon_\bold{k}|$, which is the case in most of the Brillouin Zone as  $|\Delta|$ is small compared to other band parameters. Thus the change in the RIXS intensity comes mainly from pairs where at least one quasiparticle is close to the Fermi surface, there the energy spectrum and density of states change significantly. For a pair of quasiparticles, one close to the Fermi surface, where $|\epsilon_{\bold{k}_1}|< \Delta_{\bold{k}_1}$, with $E_{\bold{k}_1} \sim |\Delta_{\bold{k}_1}|$, and $E_{\bold{k}_2} \sim \epsilon_{\bold{k}_2}$, we have $\Delta\omega \sim |\Delta_{\bold{k}_1}| + \epsilon_{\bold{k}_2}$. The same pair without the pairing term will contribute to the intensity at $\Delta\omega \sim \epsilon_{\bold{k}_2}$. 
In Fig.~\ref{fig:Map} we show the intensity as a function of $\bold{Q}$ and $\Delta \omega$, as calculated from the lowest order contribution \eqref{firstorder} for the $p+ip$ superconducting state in comparison with its normal state.
The figure shows that for small $\bold{Q}$, the intensity is enhanced, which is consistent with having an energy gap forcing larger energy transfers for two quasiparticles near the Fermi sea. \\

A yet more intriguing situation is that of a superconducting order like $p_x+p_y$ which, as opposed to the $p_x+ip_y$, exhibits nodal lines. Nodal lines are unusual but in principle allowed for p-wave systems, both for so-called unitary and non-unitary states \cite{mackenzie2003superconductivity}.
In Fig. \ref{fig:Map_pp} the RIXS intensity in the nodal and anti-nodal directions, $(1,-1)$ and $(1,1)$, respectively, are depicted for such pairing. There is a striking breaking of the symmetry between the  two directions as a result of the pairing. In the absence of pairing, the intensity in the two directions is the same. To see this, consider an electron-hole pair with momenta $(k_{x_1}, k_{y_1})$, $(k_{x_1}+q,k_{y_1}+q)$, and energies $\epsilon_1$, $\epsilon_2$, which contributes to the intensity at $\Delta\omega$ in the $(1,1)$ direction. Another electron-hole pair with 
$(k_{x_1}, -k_{y_1})$, $(k_{x_1}+q,-k_{y_1}-q)$, will have the same energies, since $\epsilon(k_x,k_y) = \epsilon(\pm k_x,\pm k_y)$, but will contribute intensity in the $(1,-1)$ direction. \\
%If a superconducting pairing is introduced, we still take the approximation that most changes are related quasiparticles near the fermi sea. Assume $\epsilon_1 \sim 0$, so $E_1 \sim |\Delta_{k_1}|$, with $\Delta_{k_1}=2i\Delta(\text{sin}k_x + \text{sin}k_y)$. From Fig.\ref{fig:Map_pp} we see the change of RIXS signal after adding a pairing is more dramatic in $(1,-1)$ direction, that means for the relevant quasiparticles near the fermi surface, the value of $|\Delta_{k_1}|$ is larger in $(1,-1)$ direction, or the momenta of the quasiparticles are more toward $(1,-1)$ direction. This can be confirmed by the Brillouin zone analysis we are about to show ({\bf \new not clear \old}).

The effect of the pairing term can be understood by looking at $\mathcal{F}_{fg}^0$ over the Brillouin zone. In \eqref{firstorder}, the summation over $\bold{r}$ gives a delta function and we can write:
\bea
\mathcal{F}_{fg}^{0} = \sum_{\bold{k}} \frac{v_{\bold{k}} u_{\bold{k+Q}}}{E_{\bold{k+Q}}+\omega+i\Gamma} 
\eea
where we took $|f \rangle = b^\dagger_{-\bold{k}} b^\dagger_{\bold{k+Q}}|g \rangle$. When the system is unpaired, $|f \rangle $ describes a particle hole pair whose momenta differ by $\bold{Q}$ and energies differ by $\Delta\omega$. 

\begin{figure}

    \begin{subfigure}{0.235\textwidth}
    \includegraphics[width=\textwidth]{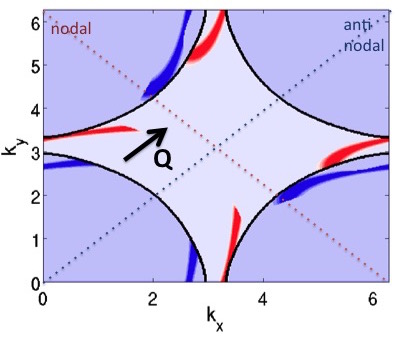}    
        \caption{}
    \end{subfigure}
    \begin{subfigure}{0.235\textwidth}
\includegraphics[width=\textwidth]{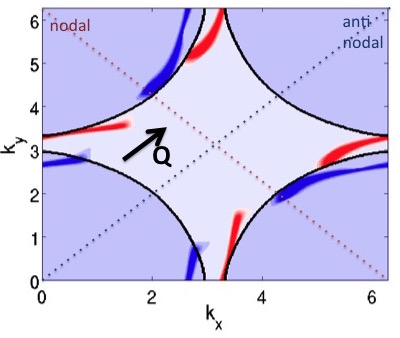}    
        \caption{}        
    \end{subfigure}

    \begin{subfigure}{0.235\textwidth}
    \includegraphics[width=\textwidth]{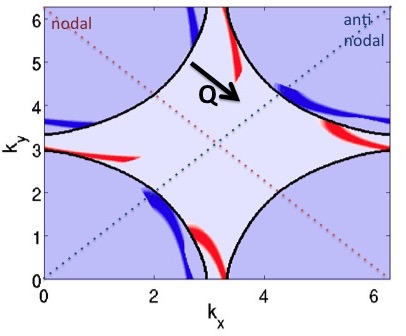}    
        \caption{}    
    \end{subfigure}
    \begin{subfigure}{0.235\textwidth}
    \includegraphics[width=\textwidth]{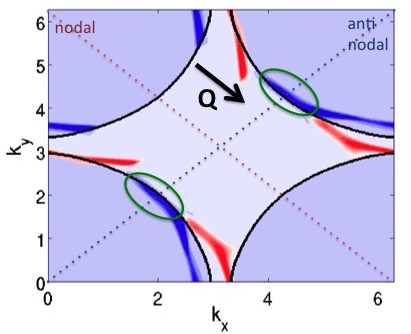}    
        \caption{}    
    \end{subfigure}

\caption{$\mathfrak{F}_{\bold{k}}$ over the Brillouin zone, for a $p_x+p_y$ pairing. The black lines show the original Fermi surface, red regions denote large values of $\mathfrak{F}_{\bold{k}}$ from electron like regions and the blue regions are the associated hole like quasiparticles. (a)  and (b): $\bold{Q}=(0.25,0.25)\pi$, anti-nodal direction, energy transfer $\Delta\omega=0.35\varepsilon$. (a): $\Delta=0$, and (b): $\Delta=0.05\varepsilon$. (c) and (d): $\bold{Q}=(0.25,-0.25)\pi$, nodal direction, energy transfer $\Delta\omega=0.35\varepsilon$. (c): $\Delta=0$, and (d): $\Delta=0.05\varepsilon$. The region marked with green  in (d) is the most affected by the pairing.}
\label{fig:Brillouin}
\end{figure}

We note that the RIXS intensity is the integral over the Brillouin zone of the function:
\bea\label{RIXSintegralBRill}
\mathfrak{F}(\bold{k}) = \frac{v_{\bold{k}} u_{\bold{k+Q}}}{E_{\bold{k+Q}}+\omega+i\Gamma} \delta(E_{\bold{k+Q}}+E_{\bold{k}}-\Delta\omega).
\eea
To identify the main contributions to the signal in momentum space we now focus on the behavior of $\mathfrak{F}(\bold{k})$. In practice, we replace the delta function by: $\delta(E) \sim e^{-(E/E_{res})^2/2}$ with $E_{res}=0.1\varepsilon$, to account for the experimental energy resolution. The result is shown in Fig.~\ref{fig:Brillouin}. Because of the symmetry of the Hamiltonian, at $\Delta=0$, $\mathfrak{F}_{\bold{k}}$ is the same at $\bold{Q}=(0.25,0.25)\pi$ and $\bold{Q}=(0.25,-0.25)\pi$, up to $90^\circ$ rotation. We can now see why for $\bold{Q}=(0.25,0.25)\pi$, in the anti-nodal direction, the effect of pairing is weaker:
$\mathfrak{F}_{\bold{k}}$ does not change a lot after turning on the pairing, since the significant regions of $\mathfrak{F}_{\bold{k}}$ are far from the line $k_x=k_y$ where the pairing, $\Delta_k=2i(\text{sin}(k_x)+\text{sin}(k_y))$ is most significant. However, in the nodal direction, $\bold{Q}=(0.25,-0.25)\pi$, a pairing term becomes much more relevant: $\mathfrak{F}_{\bold{k}}$ has significant contributions across the line $k_x=k_y$, and in those regions $\mathfrak{F}_{\bold{k}}$ is sensitive to the pairing term (noted by green circles in the plot),  resulting in a substantial change in the RIXS intensity. We thus see that the effect of pairing on intensity is sensitive to the direction of the momentum transfer, and seems to be enhanced in the nodal direction.

\section{Discussion and summary}
We have confined our discussion here to mean field BCS and made no speculation about the suitability of the treatment to strongly correlated systems and it's relevance, e. g. to high-Tc superconductors. 
At this point, it is worth mentioning, among the other probes of superconducting states, the electronic Raman scattering technique. Electronic Raman scattering essentially measures the dynamical structure factor \cite{klein1984theory}:
\bea
\tilde{S}(\bold{q},\omega) = \frac{1}{2\pi} \int_{-\infty}^\infty e^{-i\omega t}\langle \tilde{\rho}_{\bold{q}}(0)\tilde{\rho}_\bold{q}(t)\rangle,
\eea
which has a very similar form to the 4-point function measured by RIXS, and describes a similar process. In the limit where $q \xi<<1$, where $\xi$ is the coherent length, there will be a peak around $2\Delta$, which is what we get  in the small momentum limit  for RIXS using Eq. \eqref{RIXSintegralBRill}. Thus, RIXS allows for a complementary study to that of the Raman technique. It is also important to note that RIXS is especially interesting away from the BCS picture, where one can see contributions from both band structure physics and collective excitations, thus to differentiate between such effects it is of particular importance to have a well developed picture of RIXS in the absence of collective behavior. 
In particular it is of great interest to see how the present approach may affect results pertaining to the quasiparticle interpretation of RIXS in the cuprates.
Indeed, although our treatment is within a mean-field BCS picture, we remark that the method may also be of relevance to the study of cuprates. Most recent studies of RIXS in the context of cuprates have largely considered cases of insulating phases \cite{van2007theory,chen2010unraveling,vernay2008cu,Jia:2014aa}. However, RIXS experiments have been performed over a wide range of doping, including systems where itinerant electrons are present, and a description using tools developed for insulators may be insufficient. For example, in \cite{Guarise2014cuprate}, it is shown that contrary to a common interpretation, for $\text{Bi}-2212$, the magnon picture fails at a nodal direction and that a quasiparticle scenario may be an essential ingredient to understand the RIXS data there.  A different theoretical approach starts from the itinerant electrons, considering both direct \cite{benjamin2014single} and indirect RIXS processes \cite{benjamin2015probing}. It is possible to show that within this method, the RIXS signal is sensitive to particularities of the band structure \cite{kanasz2015resonant} quite far from the Fermi level, and gives results consistent with experimental studies. Another example of consideration of itinerant electrons are refs ~\cite{Igarashi2014RPA, Guarise2014cuprate} where RIXS intensity has been calculated using the random phase approximation for $Sr_2 Ir O_4$. 

In summary, we developed a general formalism to treat the RIXS intensity for a quadratic Fermi theory with arbitrary pairing. With the introduction of Majorana fermions, all quadratic Hamiltonians can be handled within the determinant method. The main formulas are summarized in  the equations (\ref{S integral},\ref{Smn},\ref{expression for Smn}) which are ready for immediate numerical use. 
Focusing on $p$-wave superconducting states, we have shown within this approach several intriguing effects on the RIXS signal. The most important findings are: a non linear shift of the RIXS absorption peak below the superconducting transition, as function of $\Delta$, and, for nodal p-wave pairing, a breaking of symmetry between the nodal and anti-nodal directions, in which, surprisingly, the effect is more pronounced in the nodal direction than the anti-nodal direction. We have seen pronounced effects of a gap scale down to a few percent of the band parameters, unfortunately, in actual $Sr_{2}RuO_{4}$, the pairing is believed to be of the order $10^{-3}\varepsilon$, and the effects discussed here will most likely be outside experimental resolution in this material with present techniques. 
However, the method introduced here, allows us to readily study other paired systems. Similar effects as described for our toy-model should be observable when carrying out RIXS measurements below and above a superconducting transition.

{\bf Acknowledgments}
It is a pleasure to thank Amit Keren, David Ellis and Marton Kanasz-Nagy for discussions.
The work of IK and YS was supported by the NSF CAREER grant DMR-0956053.  DB and ED acknowledge support from
Harvard-MIT CUA, NSF Grant No. DMR-1308435,
AFOSR Quantum Simulation MURI,
the ARO-MURI on Atomtronics, ARO MURI Quism program.

\bibliography{RIXS_References}

\section{Appendix: calculating $S^{mn}$ with pairing}
Here we give further details regarding the derivation of \eqref{expression for Smn}. Explicitly,
\begin{eqnarray}
S^{xy} & = & \langle e^{iH\tau}d_{y}e^{-iH_{y}\tau}d_{y}^{\dagger}e^{iHs}d_{x}e^{iH_{x}t}d_{x}^{\dagger}e^{-iH(t+s)}\rangle\nonumber \\
 & = & \text{{tr}[}e^{iH\tau}d_{y}e^{-iH_{y}\tau}d_{y}^{\dagger}... \nonumber \\
&...& e^{iHs}d_{x}e^{iH_{x}t}d_{x}^{\dagger}e^{-iH(t+s)-\beta H}]/\text{{tr}}[e^{-\beta H}].
\end{eqnarray}
Here, the core-holes act at sites $x$ and $y$. 

We first focus on the numerator. When replacing all the fermions with
Majorana operators, we get a combination of terms such as:
\begin{eqnarray} &
\text{{Num} }  =  \nonumber \\
 & =  \Sigma_{qmnp}\text{{tr}}[e^{iH\tau}c_{q}e^{-iH_{y}\tau}c_{m}e^{iHs}c_{n}e^{iH_{x}t}c_{p}e^{-iH(t+s)-\beta H}]\nonumber \\
 & =  \Sigma_{qmnp}\text{{tr}}[c_{q}e^{X_{4}}c_{m}e^{X_{3}}c_{n}e^{X_{2}}c_{p}e^{X_{1}}].
\end{eqnarray}
Defining $\xi x=x+N$, then the nonzero elements of $\Sigma$ are 
\begin{eqnarray*} &
\Sigma_{yyxx} = \Sigma_{\xi y,\xi y,\xi x,\xi x} =\Sigma_{\xi y,\xi y,x,x}=\Sigma_{y,y,\xi x,\xi x} =\frac{1}{16} \nonumber\\ &
\Sigma_{y,\xi y,x,\xi x} = \Sigma_{\xi y,y,\xi x,\xi x} = -\Sigma_{y,\xi y,\xi x,x} =-\Sigma_{\xi y,y,x,\xi x} = -\frac{1}{16} \nonumber\\ &
\Sigma_{y,y,x,\xi x}=\Sigma_{y,\xi y,x,x}=\Sigma_{\xi y,\xi y,\xi x,x}=\Sigma_{\xi y,y,\xi x,\xi x}=\frac{i}{16}\nonumber \\ &
\Sigma_{y,y,\xi x,x}=\Sigma_{\xi y,y,x,x}=\Sigma_{\xi y,\xi y,x,\xi x}=\Sigma_{y,\xi y,\xi x,\xi x}=-\frac{i}{16}.
\end{eqnarray*} 

Using the relation: $c_{m}e^{A_{i,j}c_{i}c_{j}}=e^{A_{i,j}c_{i}c_{j}}c_{m'}(e^{4A})_{m,m'}$
(same indices are summed over), we can move all the Majorana fermions to the right, yielding:
\bea
\text{{Num}}&=&\Sigma_{qmnp}(e^{X_{1}})_{p,p'}(e^{X_{2}}e^{X_{1}})_{n,n'}(e^{X_{3}}e^{X_{2}}e^{X_{1}})_{m,m'} \nonumber \\
&\times& \text{{tr}}[e^{Z_{ij}c_{i}c_{j}}c_{m'}c_{n'}c_{p'}c_{q}]\label{eq:Numerator}
\eea
where $e^{Z_{ij}c_{i}c_{j}}=e^{X_{4}}e^{X_{3}}e^{X_{2}}e^{X_{1}}$ . 
Now the task is to calculate traces of the form: 
\begin{eqnarray}&
{\bf T}_{mnpq}  = \text{{tr}}[e^{Z_{ij}c_{i}c_{j}}c_{m}c_{n}c_{p}c_{q}] \nonumber \\
&=\text{{tr}}[e^{Z_{ij}c_{i}c_{j}}(\delta_{mn}+\frac{{c_{m}c_{n}-c_{n}c_{m}}}{2})(\delta_{pq}+\frac{{c_{p}c_{q}-c_{q}c_{p}}}{2})]\nonumber \\
 & =  \text{{tr}}[e^{Z_{ij}c_{i}c_{j}}(\frac{{1}}{4}{\cal M}{\cal N} +\frac{{1}}{2}{\cal M} \delta_{pq}+\frac{{1}}{2}{\cal N} \delta_{nm}+\delta_{mn}\delta_{pq})],\label{eq:maintr}
\end{eqnarray}
where ${\cal M}=M_{ij}c_ic_j$. $M=|m\rangle\langle n|-|n\rangle\langle m|$, $N=|p\rangle\langle q|-|q\rangle\langle p|$.
Now that $M$ and $N$ are anti-symmetric matrices and we can write
${\cal M} =\frac{{\partial}}{\partial\alpha}e^{\alpha {\cal M}}\mid_{\alpha=0}$,
and use the trace formula \eqref{Trace formula} to calculate ${\bf T}$. First we find:
\begin{eqnarray}
&&\text{{tr}}(e^{Z_{ij}c_{i}c_{j}}\frac{{d}}{d\alpha}e^{\alpha M_{ij}c_{i}c_{j}}|_{\alpha=0})  =  \frac{{\partial}}{\partial\alpha}\text{{tr}}(e^{Z_{ij}c_{i}c_{j}}e^{\alpha M_{ij}c_{i}c_{j}})|_{\alpha=0}\nonumber \\
 & = & \frac{{1}}{2}\sqrt{\text{{det}}(1+e^{4Z}e^{4\alpha M})}\text{{tr}}[\frac{{4e^{4Z}M}}{1+e^{4Z}e^{4\alpha M}}]\mid_{\alpha=0}\nonumber \\
 & = & 2\sqrt{\text{{det}}(1+e^{4Z})}\{(1+e^{-4Z})_{nm}^{-1}-(1+e^{-4Z})_{mn}^{-1}\}
\end{eqnarray}
Next, we define $B=\frac{{1}}{1+e^{-4Z}}$ and ${\cal D}=\text{{det}}(1+e^{4Z})$. Then, 
\begin{eqnarray}
&& \frac{{\partial}}{\partial\beta}\frac{{\partial}}{\partial\alpha}\text{{tr}}(e^{Z}e^{\alpha M}e^{\beta N}) \nonumber \\
 & = &  \sqrt{{\cal D}}\{4\text{{tr}}(BM)\text{{tr}}(BN) \nonumber \\
&-& 8\text{{tr}}(BNBM)+8\text{{tr}}(BMN)\} 
\end{eqnarray} 
The last step we take $\alpha=0$, and $\beta=0$. Plugging the result
from the above two equations into Eq. \eqref{eq:maintr}, we find:
\begin{eqnarray}&
{\bf T}_{mnpq}  =  \sqrt{{\cal D}}\{(B_{nm}-B_{mn}+\delta_{mn})(B_{qp}-B_{pq}+\delta_{pq})\nonumber \\
 &   +2B_{qm}(\delta_{np}-B_{np})+2B_{pn}(\delta_{mq}-B_{mq})\nonumber \\
 &   -2B_{pm}(\delta_{nq}-B_{nq})-2B_{qn}(\delta_{mp}-B_{mp})\}
\end{eqnarray}
Noticing that since $Z$ is anti-symmetric, $B_{nm}+B_{mn}=\delta_{mn}$,
we get:
\begin{eqnarray}&
{\bf T}_{mnpq} =4\sqrt{{\cal D}}(B_{nm}B_{qp}+B_{qm}B_{pn}-B_{pm}B_{qn})  \label{Final T}
\end{eqnarray}
We plug \eqref{Final T} back into Eq.\eqref{eq:Numerator}:
\bea &
S^{xy} =\\ \nonumber & \Sigma_{qmnp}(e^{X_{1}})_{p,p'}(e^{X_{2}}e^{X_{1}})_{n,n'}(e^{X_{3}}e^{X_{2}}e^{X_{1}})_{m,m'} T_{m'n'p'q}.
\eea
We see that $S^{xy}$ is comprised of 3 terms corresponding to the terms on the right hand side of Eq. \eqref{Final T}. We first focus on the first term:
\bea &
S_1 =\\ \nonumber & 4\Sigma_{qmnp}\sqrt{{\cal D}}B_{nm}B_{qp}(e^{X_{1}})_{p,p'}(e^{X_{2}}e^{X_{1}})_{n,n'}(e^{X_{3}}e^{X_{2}}e^{X_{1}})_{m,m'}.
\eea
It will be convenient to denote $K\equiv e^{-4ih_{n}\tau}e^{4ihs}e^{4ih_{m}t}e^{4i(\tau-t-s)h}$
, and $N_{\beta}\equiv \frac{{1}}{1+e^{4\beta h}}$. With this notation we have $e^{Z}=K\frac{{N_{\beta}}}{1-N_{\beta}}$,
$B=(1+\frac{{1-N_{\beta}}}{N_{\beta}}K^{-1})^{-1}$ . And we find:
\begin{equation}
S_1=\Sigma_{qmnp}(e^{X_{3}}e^{X_{2}}e^{X_{1}}B^{T}(e^{X_{2}}e^{X_{1}})^{T})_{mn}(e^{X_{1}}B^{T})_{pq},
\end{equation}
where $T$ is the matrix transpose. In order to get convenient expressions in the low temperature
limit ($\beta\rightarrow\infty$), we have to calculate $e^{X_{1}}B(e^{X_{1}})^{T}$.
Using that for anti-symmetric matrix $h$, $e^{-h}=(e^{h})^{T}$,
we write:
\begin{eqnarray*}
e^{-4\beta h}B^{T}(e^{-4\beta h})^{T} & = & \frac{{N_{\beta}}}{1-N_{\beta}}\frac{{\frac{{1-N_{\beta}}}{N_{\beta}}K^{-1}}}{1+\frac{{1-N_{\beta}}}{N_{\beta}}K^{-1}}\frac{{1-N_{\beta}}}{N_{\beta}}\\
 & = & K^{-1}\frac{{1}}{N_{\beta}+(1-N_{\beta})K^{-1}}(1-N_{\beta}) \\
&=& \frac{{1}}{1-N_{\beta}+N_{\beta}K}(1-N_{\beta}),
\end{eqnarray*}
and
\[
e^{-4\beta h}B^{T}=\frac{{N_{\beta}}}{1-N_{\beta}}\frac{{1}}{1+K\frac{{N_{\beta}}}{1-N_{\beta}}}=N_{\beta}\frac{{1}}{1-N_{\beta}+KN_{\beta}}.
\]
Using the above results and summing over $m,n,p,q$ , we have:
\begin{eqnarray}
S_1&=&\sqrt{{\text{{det}}(F)}}(\Lambda_{y,x}+\Lambda_{\xi  y,\xi x}-i\Lambda_{\xi y,x}+i\Lambda_{y,\xi  x})\nonumber \\
&\times& (\Gamma_{y,x}+\Gamma_{\xi y,\xi x}+i\Gamma_{\xi y,x}-i\Gamma_{y,\xi x})
\end{eqnarray}
where
\begin{eqnarray}
\Lambda & = & e^{ihs}e^{ih_{x}t}e^{i(\tau-t-s)h}G^{-1}(1-N_{\beta})e^{-i(\tau-t-s)h}e^{-ih_{x}t}\nonumber \\
\Gamma & = & e^{i(\tau-t-s)h}N_{\beta}F^{-1},
\end{eqnarray}
and $F=1-N_{\beta}+KN_{\beta}$, $G=1-N_{\beta}+N_{\beta}K$.
Similarly, the second term is written as:
\bea\label{term2}
S_2 &=& \Sigma_{qmnp}(e^{X_2}e^{X_1}B^T e^{-X_1})_{pn}(e^{X_3}e^{X_2}e^{X_1}B^T)_{mq} \nonumber \\
&=&(\Lambda^{(2)}_{y,y}+\Lambda^{(2)}_{\xi y,\xi  y}+i\Lambda^{(2)}_{\xi  y,y}-i\Lambda^{(2)}_{y,\xi  y}) \nonumber \\
&\times&(\Gamma^{(2)}_{x,x}+\Gamma^{(2)}_{\xi  x,\xi  x}+i\Gamma^{(2)}_{\xi  x,x}-i\Gamma^{(2)}_{x,\xi x}),
\eea
where $\Lambda^{(2)} = e^{-ihs}\Lambda e^{ih_xt}$ and $\Gamma^{(2)}=e^{ihs}e^{ih_xt}\Gamma$.
For the third term $S_3$, 
\bea\label{term2}
S_3 &=& \Sigma_{qmnp}(e^{X_2}e^{X_1}B^T)_{nq}(e^{X_3}e^{X_2}e^{X_1}B^Te^{-X_1})_{mp} \nonumber \\
&=&(\Lambda^{(3)}_{x,y}-\Lambda^{(3)}_{\xi x,\xi y}-i\Lambda^{(3)}_{\xi x,y}-i\Lambda^{(3)}_{x,\xi y}) \nonumber \\
&\times&(\Gamma^{(3)}_{x,y}-\Gamma^{(3)}_{\xi y,\xi x}+i\Gamma^{(3)}_{\xi x,y}+i\Gamma^{(3)}_{x,\xi y})
\eea
where $\Lambda^{(3)} = \Lambda e^{ih_xt}$ and $\Gamma^{(3)}=e^{ih_xt}\Gamma$.

The terms $S_2$ and $S_3$ have a special behavior when either the core-hole potential or the superconducting pairing vanishes as follows:

(I) $S_2$ does not contribute to the inelastic signal when the core-hole potential $U_c$ is 0: In that case, $K=\mathbb{I}$, and $S_2$ only depends on $t$ and $\tau$, so $S_2$ only contributes to the elastic scattering. 

(II) $S_3$ vanishes when there is no pairing, in that case the matrices $\Lambda^{(3)}$ and $\Gamma^{(3)}$ have the special property that $\Lambda^{(3)}(x,y) = \Lambda^{(3)}(\xi x,\xi y)$, $\Lambda^{(3)}(x,\xi y) = -\Lambda^{(3)}(\xi x,y)$, so that $S_3$ vanishes.
\end{document}